\documentclass[runningheads]{llncs}
\usepackage{graphicx}
\usepackage{amsfonts}
\usepackage{amssymb}
\usepackage{amsmath}
\DeclareMathOperator{\rk}{rank}
\def\F{\widetilde{F}}

\begin{document} 
\title{Betti Curves of Rank One Symmetric Matrices\thanks{This work was supported by NIH R01 NS120581 to CC. We thank Enrique Hansen and Germán Sumbre of École Normale Supérieure for providing us with the calcium imaging data that is used in our Applications section.}}

%
%
\author{Carina Curto\inst{1} \and
Joshua Paik\inst{1} \and
Igor Rivin\inst{2}}
\authorrunning{C. Curto et al.}
%
\institute{The Pennsylvania State University
\and
Edgestream Partners LP \& Temple University}
%
\maketitle              
\begin{abstract}

Betti curves of symmetric matrices were introduced in \cite{Giusti_et_al} as a new class of matrix invariants that depend only on the relative ordering of matrix entries.  These invariants are computed using persistent homology, and can be used to detect underlying structure in biological data that may otherwise be obscured by monotone nonlinearities. Here we prove three theorems that characterize the Betti curves of rank 1 symmetric matrices. We then illustrate how these Betti curve signatures arise in natural data obtained from calcium imaging of neural activity in zebrafish. 

\end{abstract}
\section{Introduction}

Measurements in biology are often related to the underlying variables in a nonlinear fashion. For example, a brighter calcium imaging signal indicates higher neural activity, but a neuron with twice the activity of another does not produce twice the brightness. This is because the measurement is a monotone nonlinear function of the desired quantity. How can one detect meaningful structure in matrices derived from such data? One solution is to try to estimate the monotone nonlinearity, and invert it. A different approach, introduced in \cite{Giusti_et_al}, is to compute new matrix invariants that depend only on the relative ordering of matrix entries, and are thus invariant to the effects of monotone nonlinearities.

\begin{figure}[!ht]
\centering
\includegraphics[width = \textwidth]{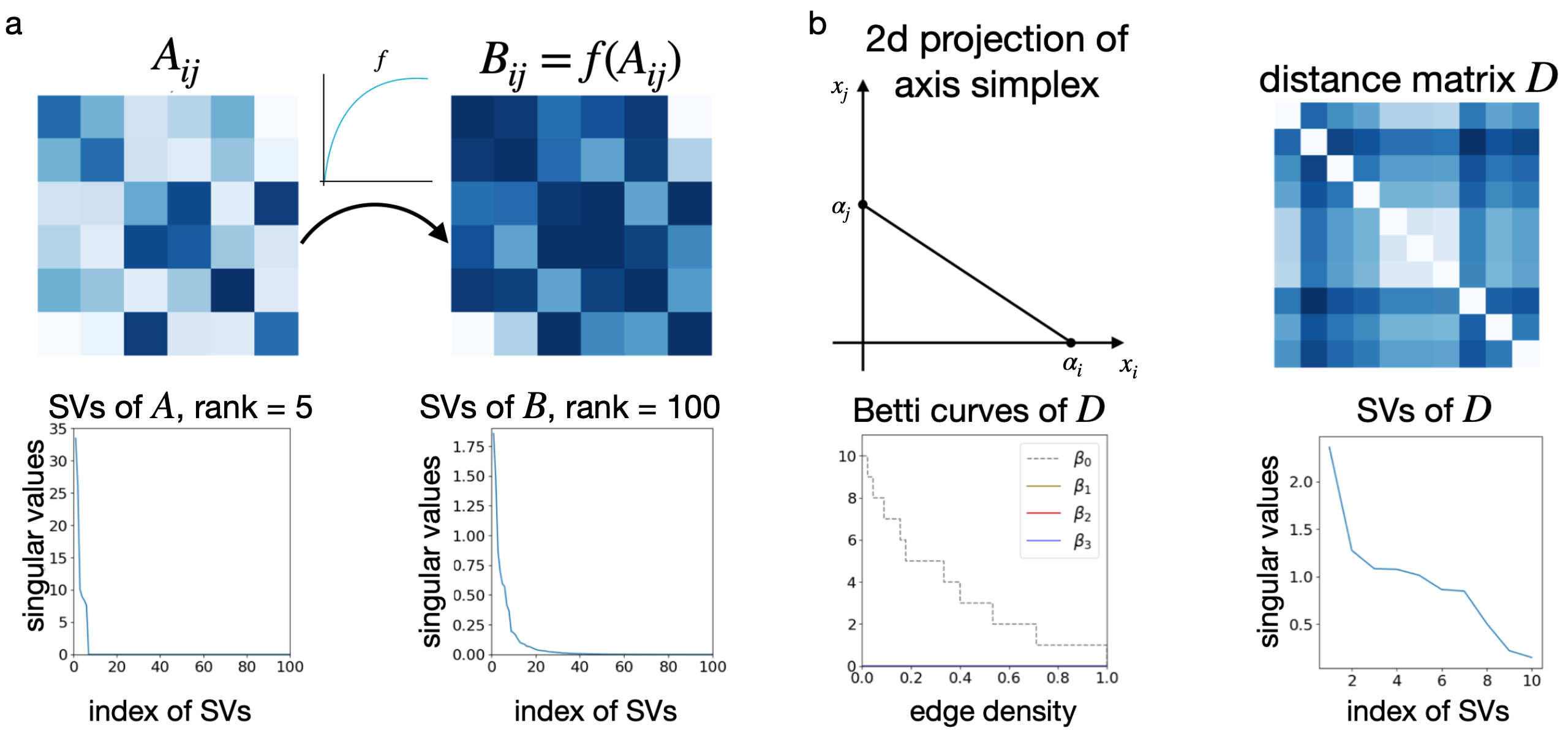}
\caption{{\bf Low rank structure is obscured by monotone nonlinearities.} The matrix $A$ is a $100 \times 100$ symmetric matrix of rank $5$. $B$ is obtained from $A$ by applying the monotone nonlinearity $f(x) = 1 - e^{-5x}$ entrywise. This alters the singular values so that $B$ appears to be full rank, but this is purely an artifact of the nonlinearity. (b) An axis simplex in $\mathbb{R}^{10}$ generates a distance matrix that is full rank, and this is evident in the singular values. The Betti curves, however, are consistent with an underlying rank of 1.}  
\label{fig:fig1}
\vspace{-.15in}
\end{figure}

Figure~\ref{fig:fig1}a illustrates the pitfalls of trying to use traditional linear algebra methods to estimate the underlying rank of a matrix in the presence of a monotone nonlinearity. The original $100\times100$ matrix $A$ is symmetric of rank 5 (top left cartoon), and this is reflected in the singular values (bottom left). In contrast, the matrix $B$ with entries $B_{ij} = f(A_{ij})$ appears to be full rank, despite having exactly the same {\it ordering} of matrix entries: $B_{ij} > B_{k\ell}$ if and only if $A_{ij} > A_{k\ell}$. The apparently high rank of $B$ is purely an artifact of the monotone nonlinearity $f$. This motivates the need for matrix invariants, like Betti curves, that will give the same answer for $A$ and $B$. Such invariants depend only on the ordering of matrix entries and do not ``see'' the nonlinearity \cite{Giusti_et_al}.

In this paper we will characterize all Betti curves that can arise from rank 1 matrices. This provides necessary conditions that must be satisfied by any matrix whose underlying rank is 1 -- that is, whose ordering is the same as that of a rank 1 matrix.  We then apply these results to calcium imaging data of neural activity in zebrafish and find that correlation matrices for cell assemblies have Betti curve signatures of rank 1.

It turns out that many interesting matrices have an underlying rank of 1. For example, consider the distance matrix induced by an \textit{axis simplex}, meaning a simplex in $\mathbb{R}^n$ whose vertices are $\alpha_1 \mathbf{e_1}, \alpha_2 \mathbf{e_2}, \dots, \alpha_n \mathbf{e_n}$, where $\mathbf{e_1}, \mathbf{e_2}, \dots, \mathbf{e_n}$ are the standard basis vectors. The Euclidean distance matrix $D$ for these points has off-diagonal entries $D_{ij} = \sqrt{\alpha_i^2 + \alpha_j^2},$ and is typically full rank (Figure~\ref{fig:fig1}b, top). But this matrix has underlying rank 1. To see this, observe that the matrix with entries $\sqrt{\alpha_i^2 + \alpha_j^2}$ has the same ordering as the matrix with entries $\alpha_i^2 + \alpha_j^2$, and this in turn has the same ordering as the matrix with entries $e^{\alpha_i^2 + \alpha_j^2} = e^{\alpha_i^2} e^{\alpha_j^2}$, which is clearly rank 1.  Although the singular values do not reflect this rank 1 structure, the Betti curves do (Figure~\ref{fig:fig1}b, bottom). 

Adding another vertex at the origin to an axis simplex yields a {\it simplex with an orthogonal corner}. The distance matrix is now $(n+1)\times(n+1)$, and is again given by $D_{ij} = \sqrt{\alpha_i^2 + \alpha_j^2}$ but with $i,j = 0,\ldots,n$ and $\alpha_0 = 0$. The same argument as above shows that this matrix also has underlying rank 1.

The organization of this paper is as follows. In Section 2 we provide background on Betti curves and prove three theorems characterizing the Betti curves of symmetric rank 1 matrices. In Section 3 we illustrate our results by computing Betti curves for pairwise correlation matrices obtained from calcium imaging data of neural activity in zebrafish. All Betti curves were computed using the well-known persistent homology package Ripser \cite{ripser}.

\section{Betti curves of symmetric rank 1 matrices}

{\bf Betti curves.} Given a real symmetric $n \times n$ matrix $M$, the ${n \choose 2}$ off-diagonal entries $M_{ij}$ for $i<j$ can be sorted in increasing order. We denote by $\widehat{M}$ be the corresponding ordering matrix, where $\widehat{M}_{ij} = k$ if $M_{ij}$ is the $k$-th smallest entry. From the ordering, we can construct an increasing sequence of graphs $\{G_t(M)\}$ for $t \in [0,1]$ as follows: for each graph, the vertex set is $1,\ldots,n$ and the edge set is 
$$E(G_t(M)) = \left\{(i,j) \mid \widehat{M}_{ij} \leq t {n \choose 2}\right\}.$$ 
Note that for $t = 0$, $G_t(M)$ has no edges, while at $t = 1$ it is the complete graph with all edges. Although $t$ is a continuous parameter, it is clear that there are only a finite number of distinct graphs in the family $\{G_t(M)\}_{t \in [0,1]}$. Each new graph differs from the previous one by the addition of an edge.\footnote{In the non-generic case of equal entries, multiple edges may be added at once.} When it is clear from the context, we will denote $G_t(M)$ as simply $G_t$.

\begin{figure}[!h]
\begin{center}
\includegraphics[width = 4.25in]{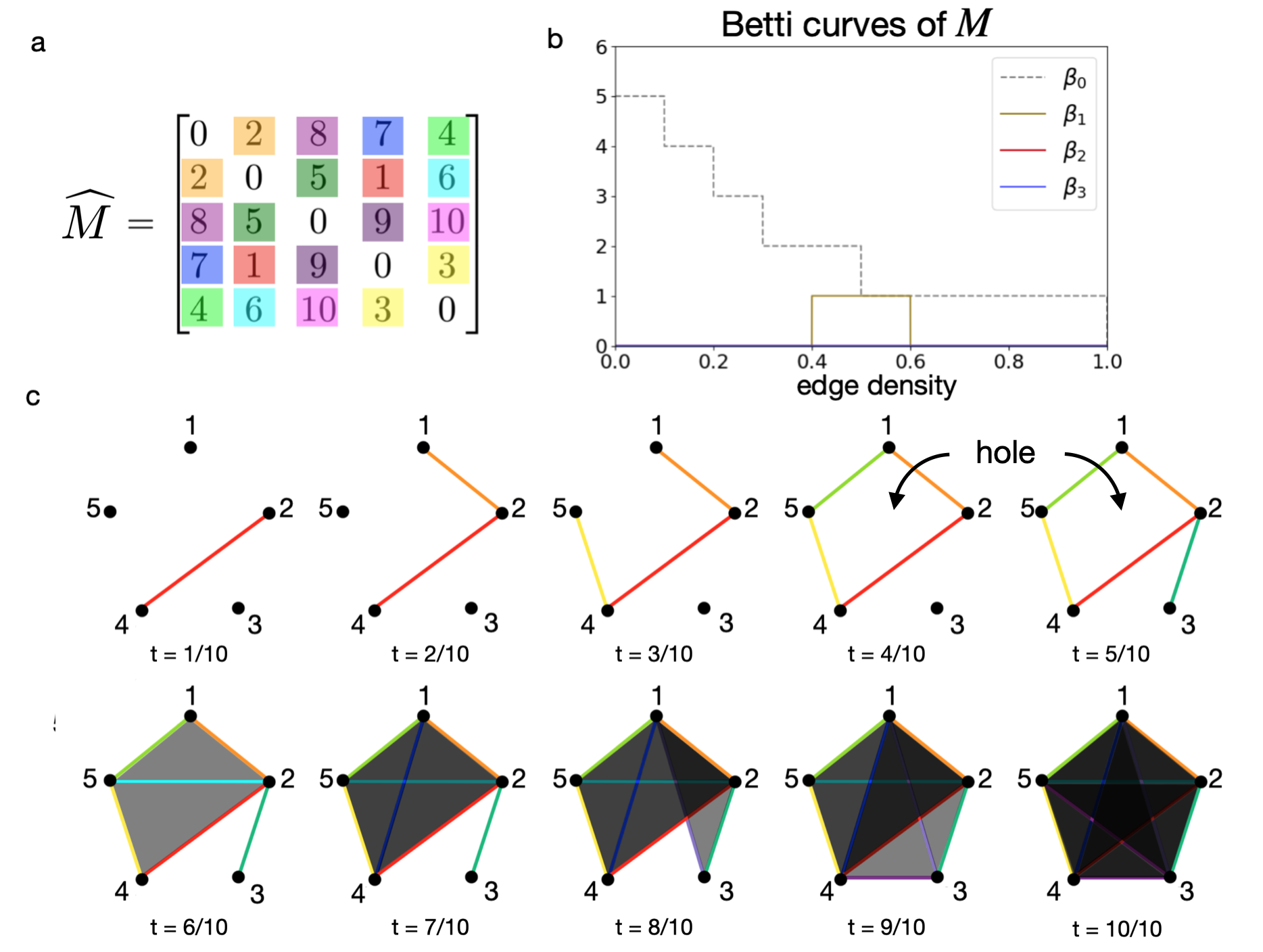}
\end{center}
\vspace{-.2in}
\caption{{\bf Betti curves of real symmetric matrices.} (a) $\widehat{M}$ is the ordering matrix for a symmetric $5 \times 5 $ matrix $M$ of rank $4$. (b) We plot the Betti curves $\beta_0, \ldots, \beta_3$ induced by $M$. (c) The filtration of clique complexes $X(G_t)$ for $\widehat{M}$ from which the Betti curves are computed. Note that at times $t = 4/10$ and $t = 5/10$, a one-dimensional hole appears, contributing to nonzero $\beta_1$ values in (b). At time $t = 6/10$, this hole is filled by the cliques created after adding edge $(2,5)$, and thus $\beta_1(t)$ goes back to zero.} 
\label{fig2}
\vspace{-.2in}
\end{figure}

For each graph $G_t$, we build a \textit{clique complex} $X(G_t)$, where every $k$-clique in $G_t$ is filled in by a $k$-simplex. We thus obtain a filtration of clique complexes $\{X(G_t) \mid t \in [0,1] \}$. The $i$-th Betti curve of $M$ is defined as
$$\beta_i(t) = \beta_i(X(G_t)) = \mathrm{rank} H_i(X(G_t),{\bf k}),$$
where $H_i$ is the $i$-th homology group with coefficients in the field ${\bf k}$. The Betti curves clearly depend only on the ordering of the entries of $M$, and are thus invariant to (increasing) monotone nonlinearities like the one shown in Figure~\ref{fig:fig1}a. They are also naturally invariant to permutations of the indices $1,\ldots,n$, provided rows and columns are permuted in the same way. See \cite{Giusti_et_al,kahle} for more details.

Figure~\ref{fig2}a shows the ordering matrix $\widehat{M}$ of a small matrix $M$. The complete sequence of clique complexes is depicted in panel c, and the corresponding Betti curves in panel b. Note that $\beta_2 = \beta_3 = 0$ for all values of the edge density $t$. On the other hand, $\beta_1(t) = 1$ for some intermediate values of $t$ where a $1$-dimensional hole arises in the clique complex $X(G_t)$. Note that $\beta_0(t)$ counts the number of connected components for each clique complex, and is thus monotonically decreasing for any matrix.

Our main results characterize Betti curves of symmetric rank 1 matrices.
We say that a vector $\mathbf{x} = (x_1, \dots, x_n)$  \textit{generates} a rank 1 matrix $M$ if $M = \mathbf{x^T}\mathbf{x}.$ Perhaps surprisingly, there are significant differences in the ordering matrices $\widehat{M}$ depending on the sign pattern of the $x_i$. The simplest case is when $M$ is generated by a vector $\mathbf{x}$ with all positive or all negative entries. When $\mathbf{x}$ has a mix of positive and negative entries, then $M$ has a block structure with two diagonal blocks of positive entries and two off-diagonal blocks of negative entries. This produces qualitatively distinct $\widehat{M}$. Even more surprising, taking $M = -\mathbf{x^T}\mathbf{x}$ qualitatively changes the structure of the ordering in a way that Betti curves can detect. This corresponds to building clique complexes from $M$ by adding edges in reverse order, from largest to smallest. Here we consider all four cases: $M = \mathbf{x^T}\mathbf{x}$ and $M = -\mathbf{x^T}\mathbf{x}$ for $\mathbf{x}$ a vector whose entries are either (i) all the same sign or (ii) have a mix of positive and negative signs. 

\begin{figure}[!ht]
\centering
\includegraphics[width = 4.25in]{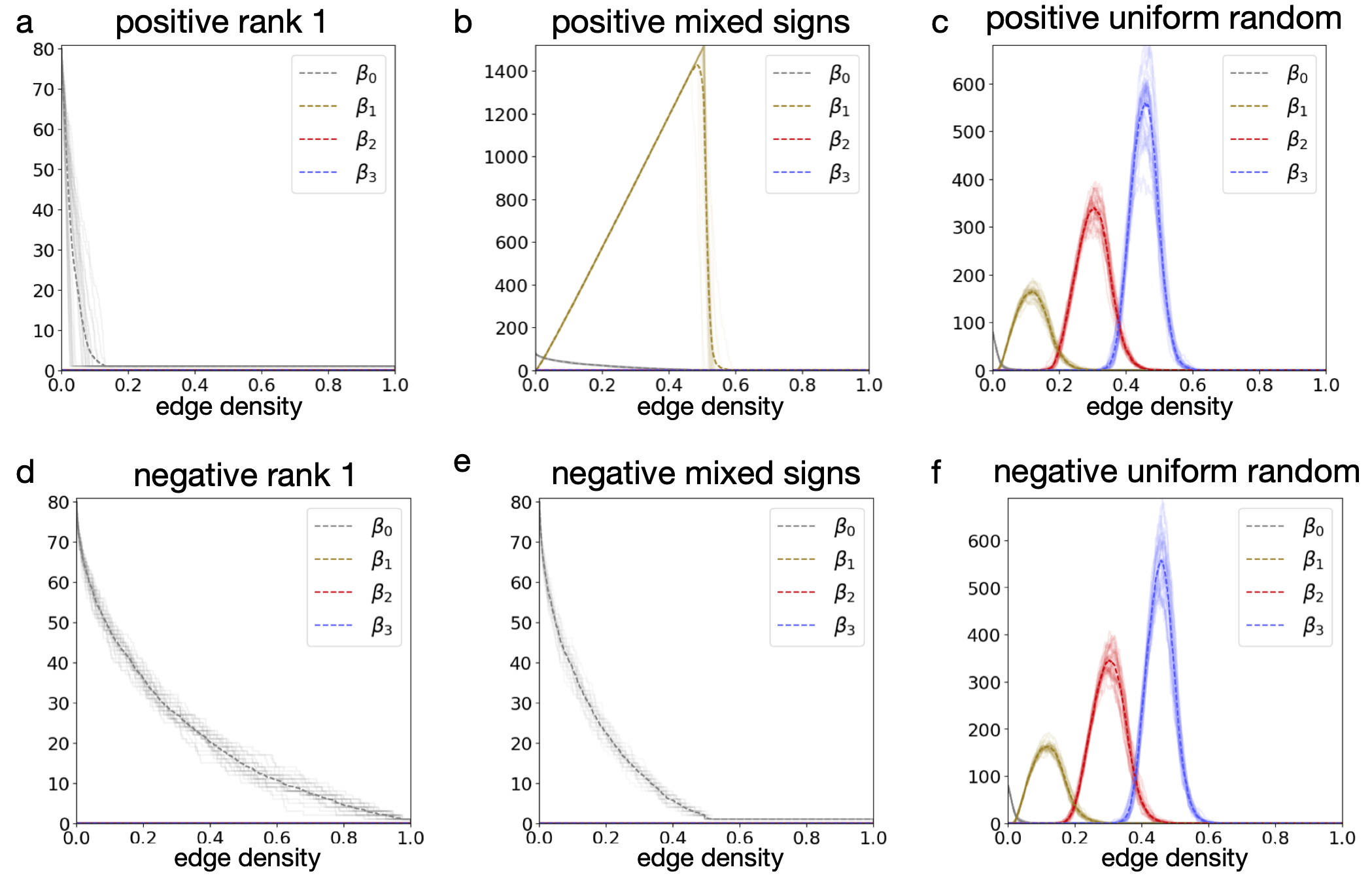}
\caption{\textbf{Betti curves for different classes of symmetric rank 1 matrices.} For each of the six figures, we generate Betti curves $\beta_0, \beta_1, \beta_2,$ and $\beta_3$ of twenty five randomly generated $80 \times 80$ symmetric matrices. (a) The Betti curve of a positive rank one matrix $M = \mathbf{x}^T\mathbf{x}$ where all entries of row vector $\mathbf{x}$ are chosen uniformly from $[0,1]$. (d) The Betti curve of a negative rank one matrix, where the only difference from the positive case is that $M =-\textbf{x}^T\textbf{x}$. (b,e) follow similar suit, except the entries of $\textbf{x}$ are chosen uniformly from the interval $[-1,1]$  and thus has mixed signs. (c,f) are random symmetric matrices constructed by making a matrix $X$ with i.i.d. entries in $[0,1]$, and setting $M_{\pm} = \pm (X + X^T)/2$. For random matrices, there is no difference in the Betti curves between the positive and negative versions.} 
\label{fig3}
\vspace{-.15in}
\end{figure}

When $\mathbf{x}$ has all entries the same sign, we say the matrix $M = \mathbf{x^T}\mathbf{x}$ is \textit{positive rank one} and $M = -\mathbf{x^T}\mathbf{x}$ is \textit{negative rank one}.
Observe in Figure~\ref{fig3}a that for a positive rank one matrix, $\beta_k(t)$ for $k > 0$ is identically zero. In Figure~\ref{fig3}d we see that the same is true for negative rank one matrices, though the $\beta_0(t)$ curve decreases more slowly than in the positive rank one case. The vanishing of the higher Betti curves in both cases can be proven.

\begin{theorem}\label{theorem_positive_rank_one}
Let $M$ be a positive rank one matrix or a negative rank one matrix. The $k$-th Betti curve $\beta_k(t)$ is identically zero for all $k >0$ and $t \in [0,1]$.
\end{theorem}

\begin{figure}
\centering
\includegraphics[width=\textwidth]{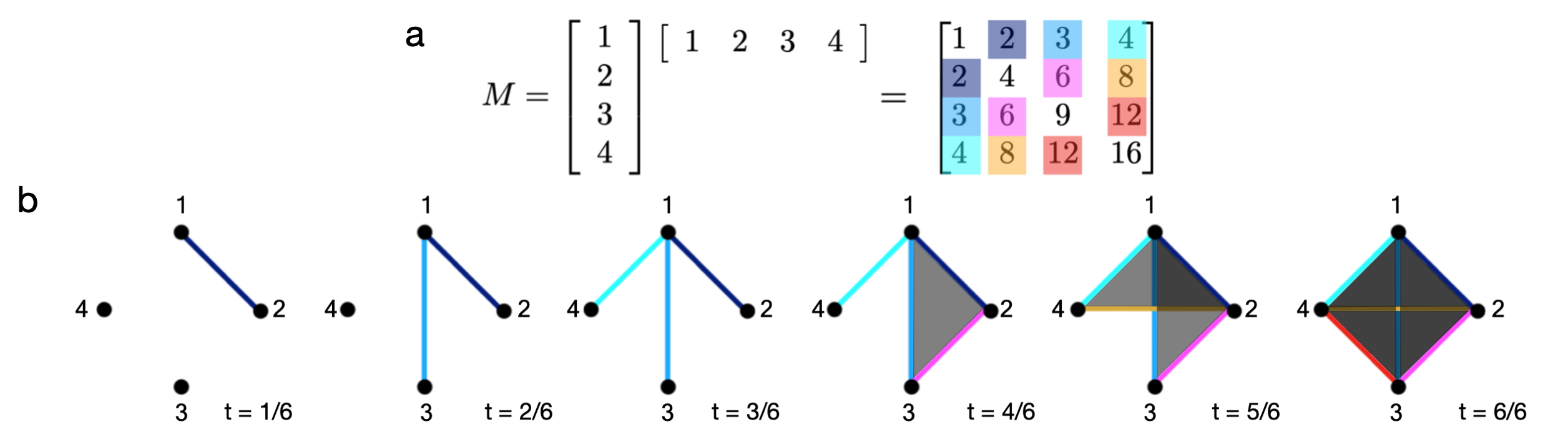}
\caption{Illustration for the proof of the positive rank one case. Note that vertex 1 has minimal value in the generating vector ${\bf x}$, and hence all other nodes are connected to 1 before any other node.}
\label{fig_pos_rank_one}
\vspace{-.2in}
\end{figure}

\begin{proof}
Without loss of generality, let $\textbf{x} = (x_1, x_2, \dots , x_n)$ be a vector such that $0 \leq x_1 \leq x_2 \leq \dots \leq x_n.$ Let $M = \textbf{x}^T \textbf{x}$ be a rank one matrix generated by $\textbf{x}$ and for a given $t \in [0,1]$, consider the graph $G_t$. It is clear that, because $x_1$ is minimal, if $(i,j)$ is an edge of $G_t$, then $(1,i)$ and $(1,j)$ are also edges of $G_t$, since $x_1x_i \leq x_ix_j$ and  $x_1x_j \leq x_ix_j$.
It follows that $X(G_t)$ is the union of a cone and a collection of isolated vertices, and hence is homotopy equivalent to a set of points. It follows that $\beta_k(t)=0$ for all $k>0$ and all $t \in [0,1]$, while $\beta_0(t)$ is the number of isolated vertices of $G_t$ plus one (for the cone component). See Figure~\ref{fig_pos_rank_one} for an example. The same argument works for $-M$, where the cone vertex corresponds to $x_n$ instead of $x_1$. $\square$
\end{proof}

The axis simplices we described in the Introduction are all positive rank one. We thus have the following corollary. 

\begin{corollary}
The k-th Betti curve of a distance matrix induced by an axis simplex or by a simplex with an orthogonal corner is identically zero for $k > 0$. 
\end{corollary}

In the cases where $\textbf{x}$ has mixed signs, the situation is a bit more complicated. Figure~\ref{fig3}b shows that for {\it positive mixed sign} matrices $M = {\bf x}^T {\bf x}$, the first Betti curve $\beta_1(t)$ ramps up linearly to a high value and then quickly crashes down to 1. In contrast, Figure~\ref{fig3}e shows that the {\it negative mixed sign} matrices $M = - {\bf x}^T {\bf x}$ have vanishing $\beta_1(t)$ and have a similar profile to the positive and negative rank one Betti curves in Figure~\ref{fig3}a,d. Note, however, that $\beta_0$ decreases more quickly than the negative rank one case, but more slowly than positive rank one. 

\begin{figure}
\centering
\includegraphics[width=\textwidth]{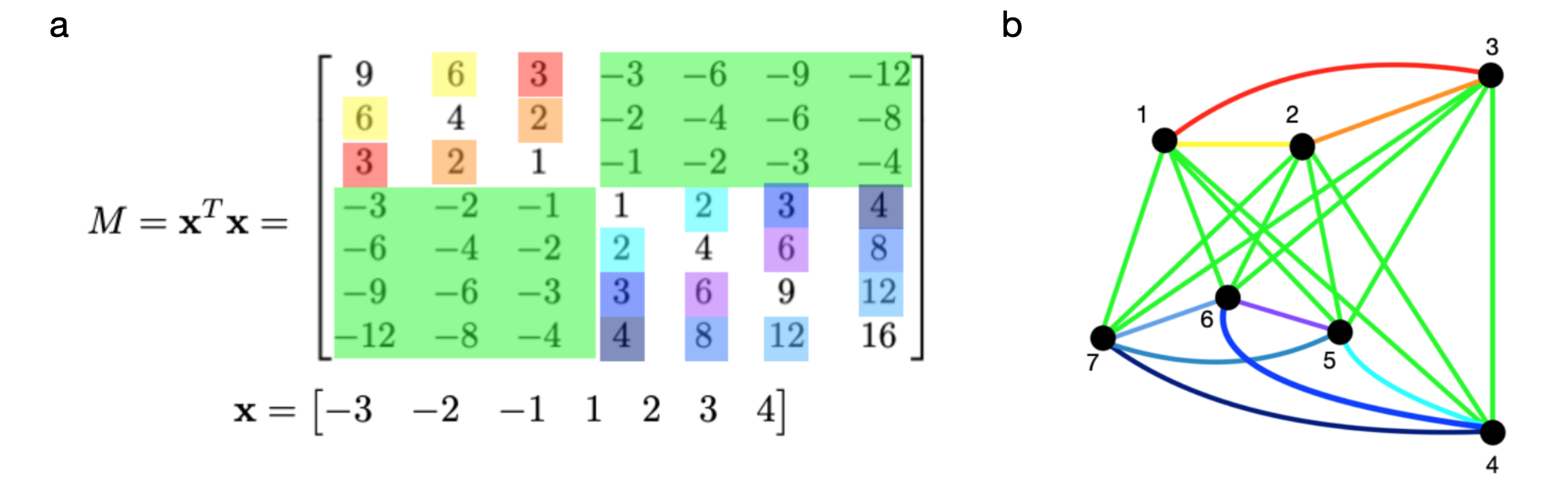}
\caption{(a) A mixed sign rank one matrix $M = {\bf x}^T {\bf x}$ with negative entries in green. (b) The graph $G_t$ induced by $M$ at $t = 1$, with edge colors corresponding to matrix colorings. The negative entries (green) get added first and form a complete bipartite graph. In the negative version, $- {\bf x}^T {\bf x}$, the green edges are added last. This leads to qualitative differences in the Betti curves.}
\label{fig:fig5}
\vspace{-.2in}
\end{figure}

Figure~\ref{fig:fig5} provides some intuition for the mixed sign cases. The matrix $M = {\bf x}^T {\bf x}$ splits into blocks, with the green edges added first. Because these edges belong to a bipartite graph, the number of 1-cycles increases until $G_t$ is a complete bipartite graph with all the green edges (see Figure~\ref{fig:fig5}b). This is what allows $\beta_1(t)$ to increase approximately linearly. Once the edges corresponding to the diagonal blocks are added, we obtain coning behavior on each side similar to what we saw in the positive rank one case. The 1-cycles created by the bipartite graph quickly disappear and the higher-order Betti curves all vanish. On the other hand, for the negative matrix $-M = -{\bf x}^T {\bf x}$, the green edges will be added last. This changes the $\beta_1(t)$ behavior dramatically, as all 1-cycles in the bipartite graph are automatically filled in with cliques because both sides of the bipartite graph are complete graphs by the time the first green edge is added.

\begin{lemma} \label{lemma:bipartite}
Let $M$ be an $n \times n$ positive mixed sign rank one matrix, generated by a vector $\mathbf{x}$ with precisely $\ell$ negative entries. Then $G_t(M)$ is bipartite for $t \leq t_0 = \ell(n-\ell)/\binom{n}{2}$ and is the complete bipartite graph, $K_{\ell,n-\ell}$, at $t = t_0$.
\end{lemma}

The next two theorems characterize the Betti curves for the positive and negative mixed sign cases. We assume $\mathbf{x} = (x_1, x_2, \dots , x_\ell, x_{\ell+1}, \dots , x_n),$ where $x_1 \leq x_2 \leq \dots \leq x_{\ell} < 0 \leq x_{\ell+1} \leq \cdots \leq x_n$.

\begin{theorem}\label{thm:positive-mixed-signs}
Let $M = \mathbf{x}^T\mathbf{x}$. Then $\beta_1(t) \leq (\ell-1)(n-\ell-1),$ with equality when $G_t$ is the complete bipartite graph $K_{\ell, n-\ell}.$ The higher Betti curves all vanish: $\beta_k(t) = 0$ for $k > 1$ and all $t \in [0,1]$.
\end{theorem}

\begin{theorem}\label{thm:negative-mixed-sign}
Let $M = -\mathbf{x}^T\mathbf{x}$. Then $\beta_k(t) = 0$ for $k > 0$ and all $t \in [0,1]$.
\end{theorem}

To prove these theorems, we need a bit more algebraic topology. Let $A$ and $B$ be simplicial complexes with disjoint vertex sets. Then \emph{the join} of $A$ and $B$, denoted $A\ast B$, is defined as 
$$A \ast B = \{ \sigma_A \cup \sigma_B \mid \sigma_A \in A \text{ and }  \sigma_B \in B\}.$$

The homology of the join of two simplicial complexes, was computed by 
J. W. Milnor \cite[Lemma 2.1]{Milnor_lemma}. We give the simpler version with field coefficients:
$$\widetilde{H}_{r+1}(A\ast B) = \bigoplus_{i+j=r} \widetilde{H}_i(A) \otimes \widetilde{H}_j(B),$$
where $\widetilde{H}$ denotes the reduced homology groups.\footnote{Note that these are the same as the usual homology groups for $i>0$.} Recall that a simplicial complex $A$ is called {\it acyclic} if $\beta_i(A) = \mathrm{rank}(H_i(A)) = 0$ for all $i>0$.

\begin{corollary}\label{cor:acyclic-join}
If $A$ and $B$ are acyclic, then $H_i(A\ast B)=0$ for all $i>1$. Also $\beta_1(A \ast B) = \rk H_1(A\ast B) = \rk(\widetilde{H}_0(A) \ast \widetilde{H}_0(B)) = (\beta_0(A)-1)(\beta_0(B)-1)$.  
\end{corollary}

\begin{example}
The complete bipartite graph, $K_{\ell, m}$, is the join of two zero dimensional complexes, $S_\ell$ and $S_m$. We see that $\beta_1(K_{\ell, m})=(\ell-1)(m-1).$ This immediately gives us the upper bound on $\beta_1(t)$ from Theorem~\ref{thm:positive-mixed-signs}.
\end{example}

\begin{example}
Let $G$ be a graph such that $V(G) = B \cup R$. Let $G_B$ be the subgraph induced by $B$ and $G_R$ the subgraph induced by $R.$ Assume that every vertex in $B$ is connected by an edge to every vertex in $R$.
Then $X(G) = X(G_B) \ast X(G_R).$
\end{example}

We are now ready to prove Theorems~\ref{thm:positive-mixed-signs} and~\ref{thm:negative-mixed-sign}. Recall that in both theorems, $M$ is generated by a vector ${\bf x}$ with positive and negative entries. We will use the notation $B = \{i \mid x_i < 0\}$ and $R = \{i \mid x_i \geq 0\}$, and refer to these as the ``negative'' and ``positive'' vertices of the graphs $G_t(M)$. Note that $|B| = \ell$ and $|R| = n - \ell$. In Figure~\ref{fig:fig5}b, $B = \{1,2,3\}$ and $R = \{4,5,6,7\}$. When $M = {\bf x}^T{\bf x}$, as in Theorem~\ref{thm:positive-mixed-signs}, the ``crossing'' edges between $B$ and $R$ are added first. (These are the green edges in Figure~\ref{fig:fig5}b.) When $M = -{\bf x}^T{\bf x}$, as in Theorem~\ref{thm:negative-mixed-sign}, the edges within the $B$ and $R$ components are added first, and the crossing edges are added last.

\begin{proof}[of Theorem~\ref{thm:positive-mixed-signs}]
We've already explained where the Betti 1 bound $\beta_1(t) \leq (\ell-1)(n-\ell-1)$ comes from (see Example 1). It remains to show that the higher Betti curves all vanish for $t \in [0,1]$. Let $t_0$ be the value at which $G_{t_0}(M)$ is the complete bipartite graph with parts $B$ and $R$. For $t \leq t_0$, when edges corresponding to negative entries of $M$ are being added, $G_t(M)$ is always a bipartite graph (see Lemma~\ref{lemma:bipartite}). It follows that $G_t(M)$ has no cliques of size greater than two, and so $X(G_t(M)) = G_t(M)$ is a one-dimensional simplicial complex. Thus, $\beta_k(t) = 0$ for all $k>1$. 

For $t > t_0$, denote by $M_B$ and $M_R$ the principal submatrices induced by the indices in $B$ and $R$, respectively. Clearly, $M_B$ and $M_R$ are positive rank one matrices, and hence by Theorem 1 the clique complexes $X(G_t(M_B))$ and $X(G_t(M_R))$ are acyclic for all $t$. Moreover, for $t > t_0$ we have that $X(G_t(M)) = X(G_t(M_B)) \ast X(G_t(M_R))$, where $\ast$ is the join (see Example 2). It follows from Corollary~\ref{cor:acyclic-join} that $\beta_k(t) = 0$ for all $k>1$. $\square$
\end{proof}

The proof of Theorem~\ref{thm:negative-mixed-sign} is a bit more subtle. In this case, the edges within each part $B$ and $R$ are added first, and the ``crossing'' edges come last. The conclusion is also different, as we prove that the $X(G_t(M))$ are all acyclic.

\begin{proof}[of Theorem~\ref{thm:negative-mixed-sign}]
Recall that $M = -{\bf x}^T{\bf x}$, and let $t_0$ be the value at which all edges for the complete graphs  $K_B$ and $K_R$ have been added, so that $G_{t_0}(M) = K_B \cup K_R$ (note that this is a disjoint union). Let $M_B$ and $M_R$ denote the principal submatrices induced by the indices in $B$ and $R$, respectively. Note that for $t \leq t_0$, $X(G_t(M)) = X(G_t(M_B)) \cup X(G_t(M_R))$. Since $M_B$ and $M_R$ are both positive rank one matrices, and the clique complexes $X(G_t(M_B))$ and $X(G_t(M_R))$ are disjoint. It follows that $X(G_t(M))$ is acyclic for $t \leq t_0$.

To see that $X(G_t(M))$ is acyclic for $t > t_0,$ we will show that $X(G_t(M))$ is the union of two contractible simplicial complexes whose intersection is also contractible.
Let $M^*$ be the positive rank one matrix generated by the vector of absolute values, $y = (|x_1|, \ldots, |x_n|)$, so that $M^* = y^T y$. Now observe that for $t > t_0$, the added edges in $G_t(M)$ correspond to entries of the matrix where $M_{ij} = - x_i x_j = |x_i| |x_j| = M^*_{ij}$. This means that the ``crossing'' edges have the same relative ordering as those of the analogous graph filtration $F_t(M^*)$. Let $s$ be the value at which the last edge added to $F_s(M^*)$ is the same as the last edge added to $G_t(M)$. (In general, $s \neq t$ and $F_s(M^*) \neq G_t(M)$.)

Since $M^*$ is positive rank one, the clique complex $X(F_s(M^*))$ is the union of a cone and some isolated vertices (see the proof of Theorem~\ref{theorem_positive_rank_one}). Let $\F_s$ be the largest connected component of $F_s(M^*)$ -- that is, the graph obtained by removing isolated vertices from $F_s(M^*)$. The clique complex $X(\F_s)$ is a cone, and hence contractible (not merely acyclic). Now define two simplicial complexes, $\Delta_1$ and $\Delta_2$, as follows:
$$\Delta_1 = X(\F_s \cup K_B) \;\; \text{and} \;\;
 \Delta_2 = X(\F_s \cup K_R).$$
It is easy to see that both $\Delta_1$ and $\Delta_2$ are contractible. Moreover, $X(G_t) = \Delta_1 \cup \Delta_2$. Since the intersection $\Delta_1 \cap \Delta_2 = X(\F_s)$ is contractible, using the Mayer-Vietoris sequence we can conclude that $X(G_t)$ is contractible for all  $t > t_0.$ $\square$
\end{proof}

\section{Application to calcium imaging data in zebrafish larvae}

Calcium imaging data for neural activity in the optic tectum of zebrafish larvae was collected by the Sumbre lab at École Normale Superieure. The individual time series of calcium activation for each neuron were preprocessed to produce neural activity rasters. This in turn was used to compute cross correlograms (CCGs) that capture the time-lagged pairwise correlations between neurons. We then integrated the CCGs over a time window of $\pm \tau_{\max}$ for $\tau_{\max} = 1$ second, to obtain pairwise correlation values:
$$C_{ij} = \frac{1}{2{\tau}_{\max}r_i r_j}\dfrac{1}{T} \int \limits_{-\tau_{\max}}^{\tau_{\max}} \int \limits_{-T}^{T} f_i(t)f_j(t + \tau) dt d\tau.$$
Here $T$ is the total time of the recording, $f_i(t)$ is the raster data for neuron $i$ taking values of $0$ or $1$ at each (discrete) time point $t$. The ``firing rate'' $r_i$ is the proportion of $1$s for neuron $i$ over the entire recording.

\begin{figure}[!ht]
    \centering
    \includegraphics[width = \textwidth]{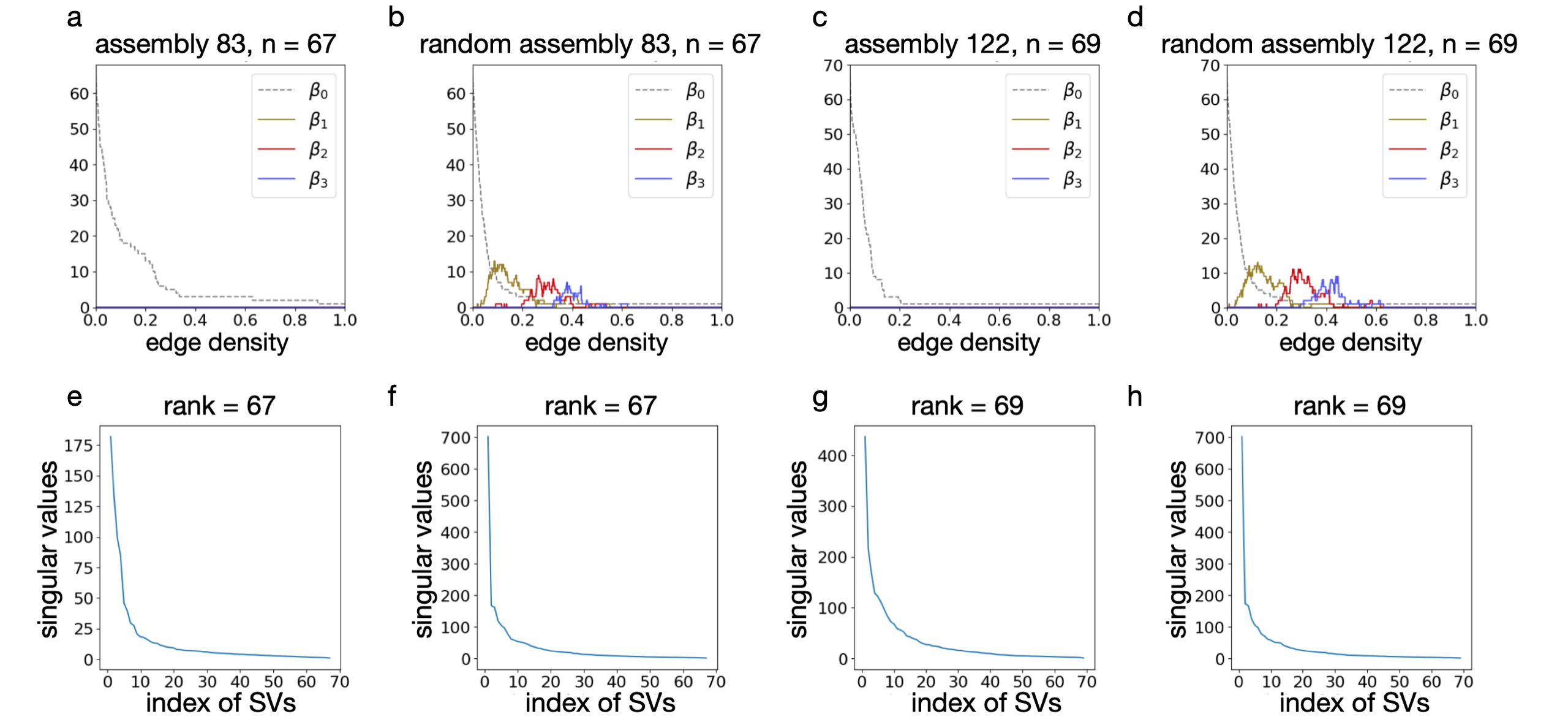}
    \caption{We plot Betti curves and singular values of two submatrices induced by preidentified cell assemblies 83 and 122 (data courtesy of the Sumbre lab at Ecole Normale Superieure). The Betti curves are consistent with a rank one matrix. In contrast, Betti curves for random assembly whose pairwise correlations come from the same CCG matrix of the same order. Below each Betti curve, we plot the normalized singular values and visually see that they are full rank and appear close to the singular values of a random assembly of the same size.}
    \label{fig6}
\end{figure}

The Sumbre lab independently identified cell assemblies: subsets of neurons that were co-active spontaneously and in response to visual stimuli, and which are believed to underlie functional units of neural computation. 
Figure~\ref{fig6} shows the Betti curves of principal submatrices of $C$ for two pre-identified cell assemblies. The Betti curves for the assemblies display the same signature we saw for rank 1 matrices, and are thus consistent with an underlying low rank. In contrast, the singular values of these same correlation matrices (centered to have zero mean) appear random and full rank, so the structure detected by the Betti curves could not have been seen with singular values. To check that this low rank Betti curve signature is not an artifact of the way the correlation matrix $C$ was computed, we compared the Betti curves to a randomly chosen assembly of the same size that includes the highest firing neuron of the original assembly. (We included the highest firing neuron because this could potentially act as a coning vertex, yielding the rank 1 Betti signature for trivial reasons.) In summary, we found that the real assemblies exhibit rank 1 Betti curves, while those of randomly-selected assemblies do not.

\section{Conclusion}

In this paper, we proved three theorems that characterized the Betti curves of rank 1 symmetric matrices. We also showed these rank 1 signatures are present in some cell assembly correlation matrices for zebrafish.
A limitation of these theorems, however, is that the converse is not in general true. Identically zero Betti curves do not imply rank 1, though our computational experiments suggest they are indicative of low rank, similar to the case of low-dimensional distance matrices \cite{Giusti_et_al}. So while we can use these results to rule out an underlying rank 1 structure, as in the random assemblies, we can not conclude from Betti curves alone that a matrix has underlying rank 1.

\subsubsection{Code}

All code used for the construction of the figures, except for Figure~\ref{fig6}, is available on Github. Code and data to produce Figure~\ref{fig6} is available upon request. 
\url{https://github.com/joshp112358/rank-one-betti-curves}.

%
%
%
%

\end{document}